%% file: manualscript.tex
\documentclass[10pt,journal,compsoc]{IEEEtran}
\usepackage{amsmath,amssymb,amsfonts,bm}
\usepackage{graphicx}
\usepackage[section]{placeins}
\usepackage{textcomp}
\usepackage{xcolor}
\usepackage{float}
\usepackage{array}
\usepackage{tabularx}
\usepackage{balance}
\usepackage{marvosym}
\usepackage{multirow}
\usepackage{booktabs}
\usepackage{algpseudocode}
\usepackage[ruled]{algorithm2e} 
\usepackage{xcolor}
\usepackage{pifont}
\usepackage{url}
\usepackage{enumitem}
\usepackage{caption}
\usepackage{subfigure}
\usepackage{bibentry}

\ifCLASSOPTIONcompsoc
  \usepackage[nocompress]{cite}
\else
  \usepackage{cite}
\fi
\ifCLASSINFOpdf
\else
\fi
\begin{document}
%
\title{Who Are the Best Adopters? \\User Selection Model for Free Trial Item Promotion}
%
%
%
%

\author{Shiqi Wang,
        Chongming Gao,
        Min Gao,~\IEEEmembership{Member,~IEEE,}
        Junliang Yu,
        Zongwei Wang,\\
        and~Hongzhi Yin,~\IEEEmembership{Senior Member,~IEEE}
\IEEEcompsocitemizethanks{\IEEEcompsocthanksitem Shiqi Wang, Min Gao, Zongwei Wang are with Chongqing University, Chongqing, China.\protect\\
E-mail: \{shiqi, gaomin, zongwei\}@cqu.edu.cn

\IEEEcompsocthanksitem Chongming Gao was with the University of Science and Technology of China, Hefei, China.\protect\\
E-mail: chongminggao@mail.ustc.edu.cn

\IEEEcompsocthanksitem Junliang Yu and Hongzhi Yin are with the University of Queensland, \\Brisbane, Australia.\protect\\
E-mail: \{jl.yu, h.yin1\}@uq.edu.au

}
\thanks{Manuscript received January **, 2022; revised ** **, 2022.\\(Corresponding author: Min Gao.)\\This work has been submitted to the IEEE for possible publication. Copyright may be transferred without notice, after which this version may no longer be accessible.
}}

%
%

\markboth{Journal of \LaTeX\ Class Files,~Vol.~**, No.~**, August~2022}%
{Shell \MakeLowercase{\textit{et al.}}: Bare Demo of IEEEtran.cls for Computer Society Journals}
%



\IEEEtitleabstractindextext{%
\begin{abstract}

With the increasingly fierce market competition, offering a free trial has become a potent stimuli strategy to promote products and attract users. By providing users with opportunities to experience goods without charge, a free trial makes adopters know more about products and thus encourages their willingness to buy. However, as the critical point in the promotion process, finding the proper adopters is rarely explored. Empirically winnowing users by their static demographic attributes is feasible but less effective, neglecting their personalized preferences.

To dynamically match the products with the best adopters, in this work, we propose a novel free trial user selection model named SMILE, which is based on reinforcement learning (RL) where an agent actively selects specific adopters aiming to maximize the profit after free trials. Specifically, we design a tree structure to reformulate the action space, which allows us to select adopters from massive user space efficiently.

The experimental analysis on three datasets demonstrates the proposed model's superiority and elucidates why reinforcement learning and tree structure can improve performance. Our study demonstrates technical feasibility for constructing a more robust and intelligent user selection model and guides for investigating more marketing promotion strategies. 
\end{abstract}

}

\maketitle

\input{LaTeX/01introduction}
\input{LaTeX/02relatedwork}

\input{LaTeX/03method}

\input{LaTeX/04experiment}
\input{LaTeX/05conclusion}
\input{LaTeX/06ack}

\IEEEdisplaynontitleabstractindextext

%
\IEEEpeerreviewmaketitle

\ifCLASSOPTIONcaptionsoff
  \newpage
\fi

\bibliographystyle{IEEEtran}
\bibliography{IEEEtran}

	\begin{IEEEbiography}[{\includegraphics[width=1in,height=1.25in,clip,keepaspectratio]{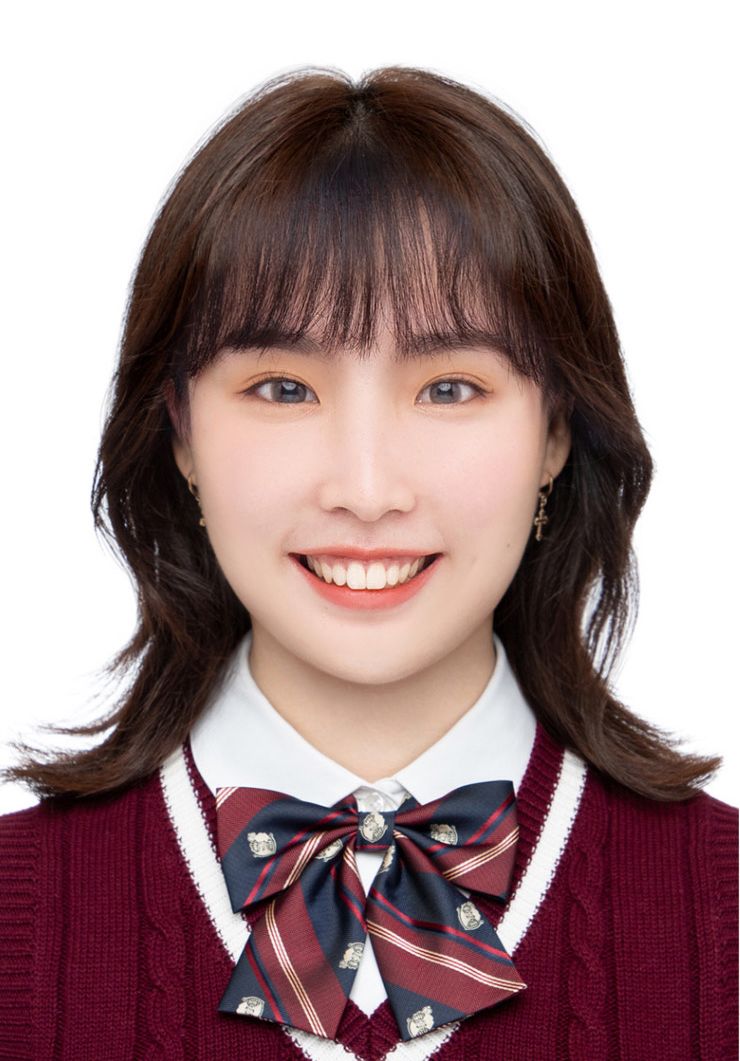}}]{Shiqi Wang}
		received the B.S. degree in Software Engineering from Chongqing University in 2020. Currently, she is a M.S. student in School of Big Data and Software Engineering at Chongqing University, Chongqing, China. Her research interests include recommender systems and reinforcement learning.
	\end{IEEEbiography}

	\begin{IEEEbiography}[{\includegraphics[width=1in,height=1.25in,clip,keepaspectratio]{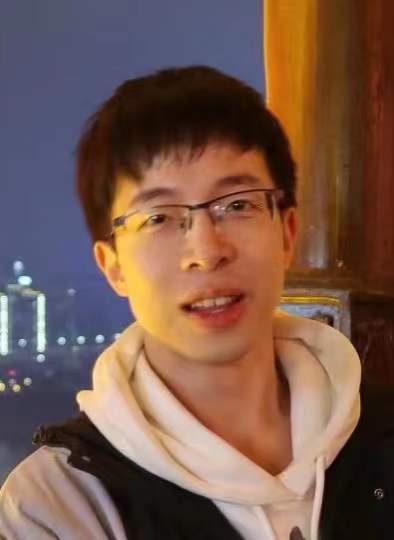}}]{Chongming Gao}
		received the B.S. and M.S. degrees in the University of Electronic Science and Technology of China (UESTC) in 2016 and 2019 respectively. Currently, he is a Ph.D. student in the University of Science and Technology of China (USTC). His research interests include recommender systems,conversational recommender system and natural language processing.
	\end{IEEEbiography}
	
	\begin{IEEEbiography}[{\includegraphics[width=1in,height=1.25in,clip,keepaspectratio]{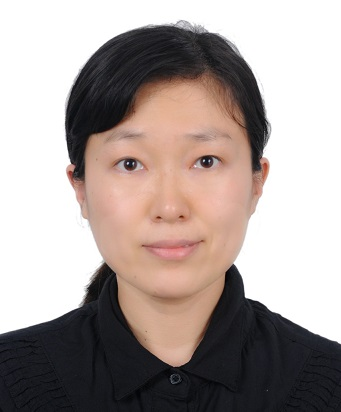}}]{Min Gao}
		received the M.S. and Ph.D. degrees in computer science from Chongqing University in 2005 and 2010 respectively. She is an associate professor at the School of Big Data \& Software Engineering, Chongqing University. Her research interests include recommendation systems, social computing, and data mining. 
	\end{IEEEbiography}

	\begin{IEEEbiography}[{\includegraphics[width=1in,height=1.25in,clip,keepaspectratio]{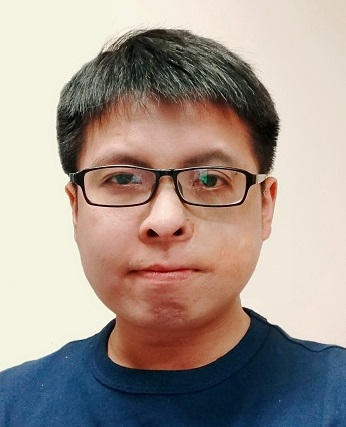}}]{Junliang Yu}
		received the B.S. and M.S. degrees in Software Engineering from Chongqing University, Chongqing, China. Currently, he is a Ph.D. student with the School of Information Technology and Electrical Engineering at the University of Queensland, Queensland, Australia. His research interests include recommender systems, social media analytics, deep learning on graphs, and self-supervised learning.
	\end{IEEEbiography}
 	\vspace{-80 mm}

	\begin{IEEEbiography}[{\includegraphics[width=1in,height=1.25in,clip,keepaspectratio]{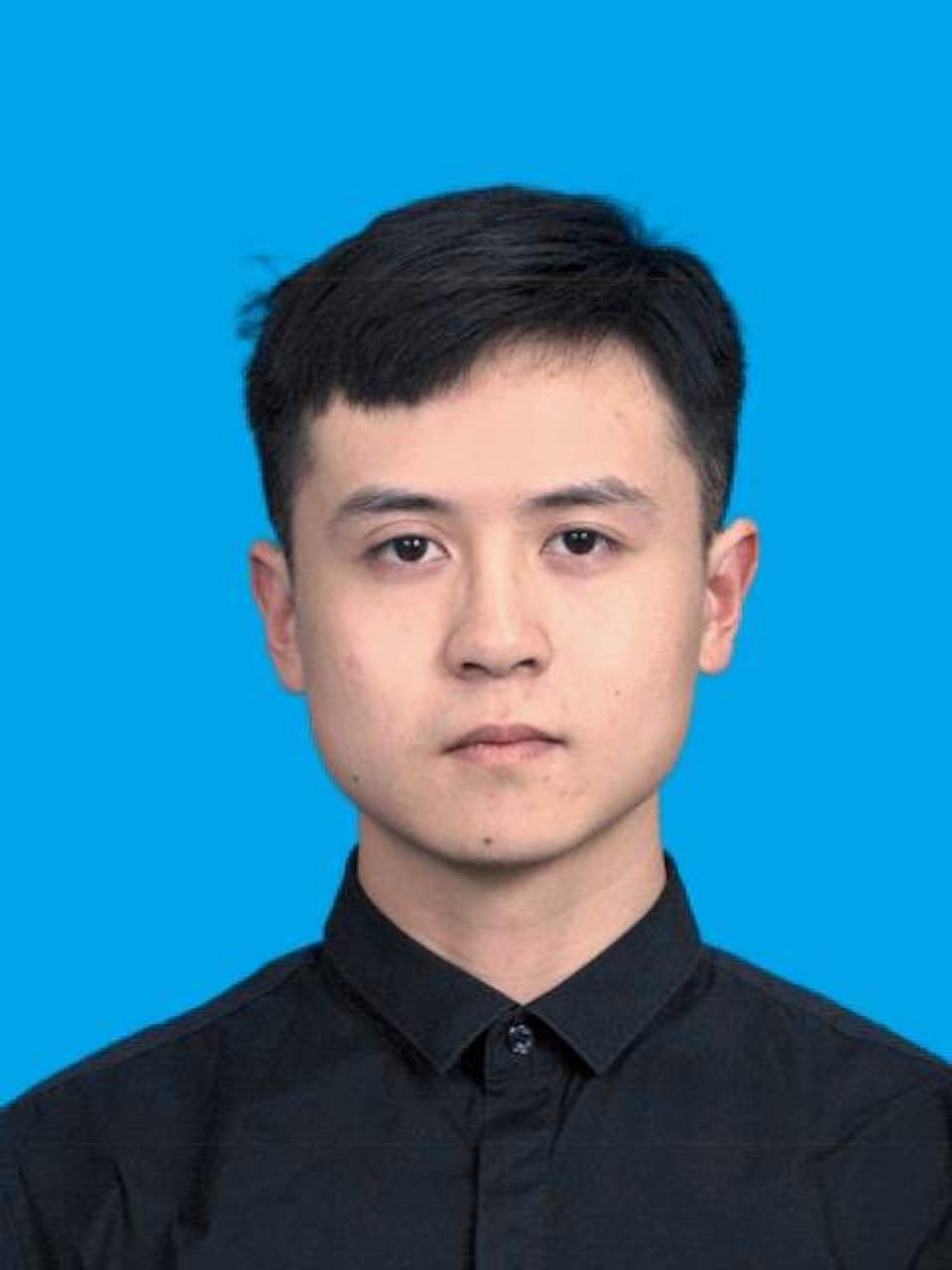}}]{Zongwei Wang}
		received the B.S. degree in Software Engineering and M.S. degree in Vehicle Engineering from Chongqing University, Chongqing, China. Currently, he works in China Securities Depository and Clearing Corporation Limited, Shanghai, China. His research interests include recommender systems and adversarial attack.
	\end{IEEEbiography}
	\vspace{-80 mm}
	
	\begin{IEEEbiography}[{\includegraphics[width=1in,height=1.25in,clip,keepaspectratio]{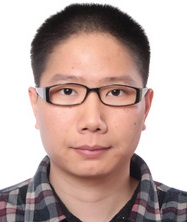}}]{Hongzhi Yin}
		received the Ph.D. degree in Computer Science from Peking University, in 2014. Currently, he works as ARC Future Fellow and associate professor with the University of Queensland, Australia. He has won 6 Best Paper Awards such as ICDE’19 Best Paper Award, DASFAA’20 Best Student Paper Award, and ACM Computing Reviews’ 21st Annual Best of Computing Notable Books and Articles as well as one invited paper in the special issue of KAIS on the best papers of ICDM 2018. His research interests include recommender system, graph embedding and mining, chatbots, social media analytics and mining, edge machine learning, trustworthy machine learning, decentralized and federated learning, and smart healthcare.
	\end{IEEEbiography}

\end{document}

%% file: LaTeX/01introduction.tex
\section{Introduction}
The development of mobile technology and fierce market competition has promoted the vigorous development of online platforms. Varieties of E-commerce services such as video/music streaming platforms now play a crucial role in our daily lives. However, with the rise of online items and the limitation of platform display pages, significant exposure opportunities only concentrate on a few popular items.
\begin{figure}[!t]
\includegraphics[width=0.95\linewidth]{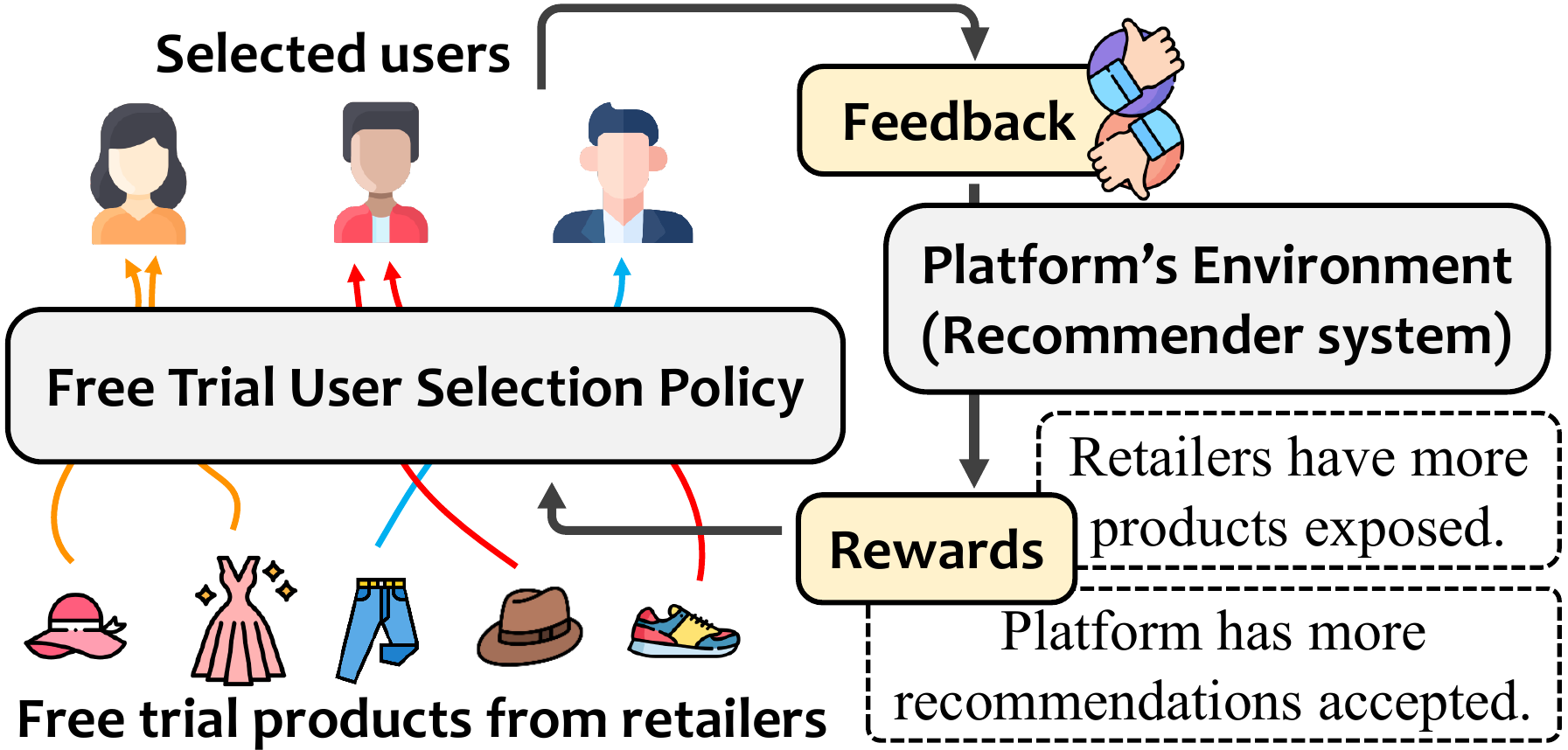}
\centering
\caption{Pipeline of free trial promotion.}
\label{fig:intro}
\vspace{-0.3cm}
\end{figure}
Compared with popular items, low-exposure items usually hold a more flexible pricing strategy due to less similar competitive products, thus embracing relatively large marginal profit \cite{DBLP:journals/pvldb/YinCLYC12}. Meanwhile, they are more likely to surprise users and thus increase their loyalty and satisfaction to the platform. Its cumulative benefits often exceed expectations \cite{li2019both}. For instance, more than a quarter of Amazon’s book sales come from outside its top 100,000 titles \cite{anderson2006long}. 
Reasons for the unbalanced exposure opportunities lie in the recommender system behind the page display mechanism. Generally speaking, recommendation algorithms are usually based on collaborative filtering \cite{9246261}, which filters items of user interest based on user/item similarity. Owing to various similar items and rich transactions, popular items are more inclined to be recommended by algorithms. However, these results will undoubtedly reduce the diversity and coverage of items, exacerbate the cold-start problem, and further intensify the popularity bias problem in recommender system \cite{abdollahpouri2019unfairness}. As a consequence, for customers, it may lead to missing high-quality products, thereby affecting their experience and satisfaction. For platforms, it may cause a considerable loss of potential profit and a reduction of competitiveness.

To stand out from the fierce market competition, platforms often adopt various promotion methods to quickly grab customers, increase item exposure, and obtain more transactions. Typically, personnel promoting  \cite{julian1994role,glassman1992integrating}, pricing strategy  \cite{shankar2004empirical,lee1997vertical}, advertising  \cite{singh2005capital,chiu2006marketing}, and celebrity endorsement  \cite{erdogan1999celebrity,khatri2006celebrity} are commonly used to attract customers' attention. However, these approaches have problems like costly, time-consuming, and lacking user feedback. For example, adverting on the Taobao homepage is not only unable to obtain users' immediate comment on the product but also expensive (more than 20,000 US dollars per day).

To fill this gap and win brand reputation, massive companies launch free trial activities, such as Netflix for online movies and TV shows, King’s Candy Crush Saga for online games, and Kindle unlimited for eBook. Intuitively, providing the experience opportunities without charge allows customers to obtain direct sensory contact \cite{kempf1998consumer}, and can further effectively ease users’ uncertainty about the utility and quality of the chargeable products. Specially, the real feedback from adopters is helpful for improving item quality and also advantageous to pricing decisions \cite{jiao2020understanding,cheng2012optimal,niu2019pricing}.
Figure~\ref{fig:intro} illustrates the pipeline of free trial promotion. Retailers first carefully select a set of users as adopters according to the selection policy. Next, the promoted products are sent to them for free. In return, customers are obliged to give immediate feedback describing their feelings or give amendatory suggestions. Then the platform environment will return rewards signal indicating the increment of exposure which helps adjust the selection policy. In this way, retailers have more products exposed, and the platform has more recommendations accepted. 

Although free trial activities have been widely used in the practical marketing promotion scene, far too little attention has been paid to the user selection policy, which is vital in the whole process. Winnowing adopters indiscriminately or simply considering their naive sociological attributes are highly feasible but less effective. Lacking systematically analysis and guidance, these rigid and fixed approaches ignore the dynamic interaction between users and platforms, making it difficult to adapt to the flexible and changeable reality scenario. Additionally, these simple handcrafted rules are insufficient in personalization that is highly required under specific environments. 
Consequently, promoting items by free trial is not an easy case due to the following challenges: (1) systematically formalizing the loop process: ``selecting adopters $\rightarrow$ receiving feedback $\rightarrow$ train model $\rightarrow$ selecting adopters $\rightarrow$ $\dots$'' and (2) selecting appropriate users who can maximize the exposure of the item under protean interactive environment. 

Recently, reinforcement learning (RL) \cite{sutton2018reinforcement} has achieved remarkable success in scenarios requiring dynamic interaction and long-run planning, such as playing games \cite{mnih2015human,silver2016mastering}, regulating ad bidding \cite{cai2017real,jin2018real}, and dynamic resource allocation \cite{9072494,9078855}. Considering the dynamic nature of real-world promotion process, we propose \textbf{SMILE} (short for "user \textbf{S}election \textbf{M}odel w\textbf{I}th po\textbf{L}icy gradi\textbf{E}nt") framework under reinforcement learning structure to learn effective selection strategies.
It repeatedly selects trial users and improves its own selection strategies through available reward signals until the model converges. Specifically, we model the selection process as an MDP and adopt policy gradient \cite{sutton2000policy}, a well-known RL method, to learn how to make decisions for maximizing long-run rewards. For the state representation $\mathbf{s}_t$ at time $t$, we use the recurrent neural network to embed the historical actions and the corresponding rewards into a low-dimensional hidden vector. For the reward signal, we consider the observable number of Page View (PV) \cite{li2019fair,garcin2014offline} on a pre-defined target item set. Furthermore, to overcome the low convergence problem on a large action space in reinforcement learning, we further reformulate the action space through a balanced hierarchical clustering tree.

In summary, the main contributions of this paper are as follows:
\begin{itemize}
    \item To the best of our knowledge, this is the first work aiming for increasing promoted item exposure by selecting the best free trial adopters. 
    \item We formulate the problem of user selection and propose an RL-based approach to deal with the dynamic interactions between adopters and the recommender system. For more efficient selection, we design a balanced hierarchical clustering tree to reformulate the action space.
    \item We conduct extensive experiments on three public datasets to show superior performance and significant efficiency improvement of the proposed SMILE framework and elucidate why RL and the clustering tree structure can improve the performance.
\end{itemize}

%% file: LaTeX/02relatedwork.tex
\section{Related Work}

\subsection{Free Trial Marketing Strategy}
Product trial was firstly defined as a consumer's first usage experience with a brand or product by Kempf and Smith \cite{kempf1998consumer}. It gradually becomes a widely applied marketing strategy for attracting users. According to the survey of marketing week \cite{bawa2004effects}, product trials can deepen customers' brand awareness and improve product brand recognition. About 63\% of 
them tend to buy tried products.

Some studies investigated the influence between free trials and users' purchasing intention. Zhu \emph{et al.} \cite{zhu2014investigating} explore consumer intention towards free trials of technology-based services and find that perceived usefulness, perceived ease of use, perceived risk, and social influence are essential determinant factors.
Sun \emph{et al.} \cite{sun2017can} find that most users have little knowledge about new services and thus have low intention to purchase them. Free trial is an effective marketing method to improve users' beliefs about the service. 
Wang \emph{et al.} \cite{wang2013user} prove that users' experience on the mobile newspaper software after the free trial is different from before. 
Halbheer \emph{et al.} \cite{halbheer2014choosing} show that free trial strategy influences user's expectations of product quality which is closely linked to the user demand. 
Foubert and Gijsbrechts \cite{foubert2016try} find that free trial is more effective in conveying information than advertising because actual usage can quickly reduce user uncertainty. 
These studies indicate that free trials can improve users' service experience and further influence users' purchase decisions.

Some other studies concentrate on when the firm should adopt the free trial strategy. Cheng \emph{et al.} \cite{cheng2010free} find that under a strong network effect, the firm is better off offering free trial than segmenting the market by charging a price for a lower quality product. Niu \emph{et al.} \cite{niu2019pricing} find that customers' prior belief plays a key role, and the firm offers free trial only when customers' initial belief is less than a threshold. 

These studies indicate that free trials can influence users' purchase decisions and uncover the conditions under which firms should introduce the free trial product. However, there has been little discussion about the process of selecting trial objects which is significant for improving trial quality and better marketing.

\subsection{RL-based Recommendation}
Reinforcement learning (RL) has been introduced into recommender systems as its advantage of considering users' long-term feedbacks \cite{zhao2018recommendations,zou2020pseudo}. 
Zou \emph{et al.}  \cite{zou2019marlrank} formulate the ranking process as a multi-agent Markov Decision Process, where mutual interactions are incorporated to compute the ranking list.
In the 1900s, WebWatcher \cite{joachims1997webwatcher} models the web page recommendation problem as an RL problem and adopts Q-learning to improve its performance. 
Later, with the development of deep learning, combining deep learning with traditional RL methods is becoming increasingly popular in RS. DQN has been used in clinical applications, such as optimizing heparin dosage recommendations \cite{nemati2016optimal} and optimizing dosage recommendations for sepsis treatment \cite{raghu2017deep}. 
Recently, many interesting applications have emerged. Google utilizes RL to recommend more suitable video content to its users on YouTube \cite{chen2019top}. Fotopoulou \emph{et al.} \cite{fotopoulou2020interactive} design an RL-like framework for an activity recommender for students' social-emotional learning. 
Liu \emph{et al.}  \cite{liu2018towards} use RL to recommend learning activities in a class by monitoring students' learning status.

However, most RL-based models fail to serve for recommender system those need to operate on the large discrete action space. 
For DQN-based algorithm \cite{zhao2018recommendations, DBLP:conf/www/ZhengZZXY0L18} which needs to find an appropriate action from large action space by value function $\mathrm{Q}(s,a)$, to maximize the discounted cumulative reward. 
The same problem applies to DDPG-based algorithm \cite{zhao2018deep, hu2018reinforcement}; it needs to learn a specific ranking function whose complexity of sampling an action grows linearly concerning the size of the action set.
Dulac-Arnold \emph{et al.} \cite{dulac2015deep} focus on the large action space problem by modeling the state in the same continuous item embedding space and selecting the items via nearest neighborhood search.
Chen \emph{et al.}  \cite{chen2019large} propose TPGR which aims to represent the item space in the form of a balanced tree and learn a strategy, using policy networks, to select the best child nodes for every non-leaf node.
In 2021, Chen \emph{et al.} \cite{chen2021user} present a general framework to augment the training of model-free RL agents with modeling user response auxiliary tasks to improve sample efficiency and conduct experiments on industrial recommendation platforms serving billions of users to verify its benefit.

In this paper, based on the structure of TPGR, we propose a balanced tree structure SMILE to select appropriate adopters in large discrete action space to interact with the flexible and changeable scenarios.

%% file: LaTeX/03method.tex
\section{Problem Formulation}
In this section, we firstly systematically formalize a new problem called \emph{selecting the best adopters for free trial item promotion}. Then we present our approach based on reinforcement learning to solve this problem. 

Given a promoted item set $I_p$ provided for free, we aim to select $n$ adopters that can benefit the retailers and the service provider maximally, i.e., both the profit of the seller and the acceptance of the shopping platform can be improved to the most extent after free trial. We quantify this effect by maximizing the exposure of promoted items without reducing the recommender performance. Our user selection policy will be ceaselessly trained based on the received reward $r$ every round to make better decisions. The complete process is illustrated in Fig.~\ref{fig:intro}.

The whole selection process is formulated as a Markov Decision Process (MDP) in reinforcement learning \cite{sutton2018reinforcement}, whose key components are summarized as follows:
\begin{itemize}

\item \textbf{State}. The model maintains a state $\mathbf{s}_t \in \mathbb{R}^{d_s}$ at time $t$ is regarded as a vector representing information of historical interactions between promoted item $p_{i}\in I_p$ and system prior to $t$. In this paper, we obtain the $\mathbf{s}_t$ via a recurrent neural network (RNN).

\item \textbf{Action}. The model makes an action $a_t$ at time $t$ is to select one adopter for item $p_{i}$. Let $\mathbf{e}_{a_t} \in \mathbb{R}^{d_a}$ denote the representation vector of action $a_t$. In this paper, each action $a_t$ selects only one user $u$. Hence, we have $\mathbf{e}_{a_t} = \mathbf{e}_u$.

\item \textbf{Reward}. The recommender system returns a reward score $r_t$ reflecting the exposure of the target items at time $t$.

\item \textbf{Transition}. The transition function gives the next state $\mathbf{s}_{t+1}$ after taking action $a_t$. Due to the state reflecting the historical interactions of the target item, it will be changed given the newly selected user and its corresponding rewards.

\item \textbf{Policy Network}. The policy network $\pi_\theta = \pi_\theta(a_t|\mathbf{s}_t)$ is the target policy that decides how to make an action $a_t$ conditioned on the state $\mathbf{s}_t$. In this paper, policy network is designed with a fully-connected neural network and a softmax activation function on the output layer. 
It takes the state $\mathbf{s}_t$ as input and outputs a probability distribution over possible outputs. The probability of a choice $a_t$ is computed as follows:
\begin{equation}
\label{softmax}
\pi_\theta(a_t|\mathbf{s}_t)=\operatorname{Softmax}(\sigma \left(\mathbf{W_s}^{T} \mathbf{s_t}+b\right)),
\end{equation}
where $\sigma$ is a non-linear activation function, $\mathbf{W_s}$ denotes the weight matrix and $b$ is bias value. 
\end{itemize}

As a consequence, we regard a free trial user selection process as 
$(\mathbf{s}_1, a_1, r_1, \mathbf{s}_2, \cdots, \mathbf{s}_n, a_n, r_n, \mathbf{s}_{n+1})$, which represents one episode. In detail, the state vector $\mathbf{s}_t$ enters the policy network and outputs an action $a_t$, i.e., the generated next adopter; then the recommender system returns a reward $r_t$ measuring the quality of this action. The selection process will terminate at a specific state $\mathbf{s}_{n+1}$ when the pre-defined episode length is satisfied. Without loss of generality, we set the length of an episode $n$ to a fixed number \cite{cai2017real}.

\section{Proposed Method}
\begin{figure}[t]
\includegraphics[width=1\linewidth]{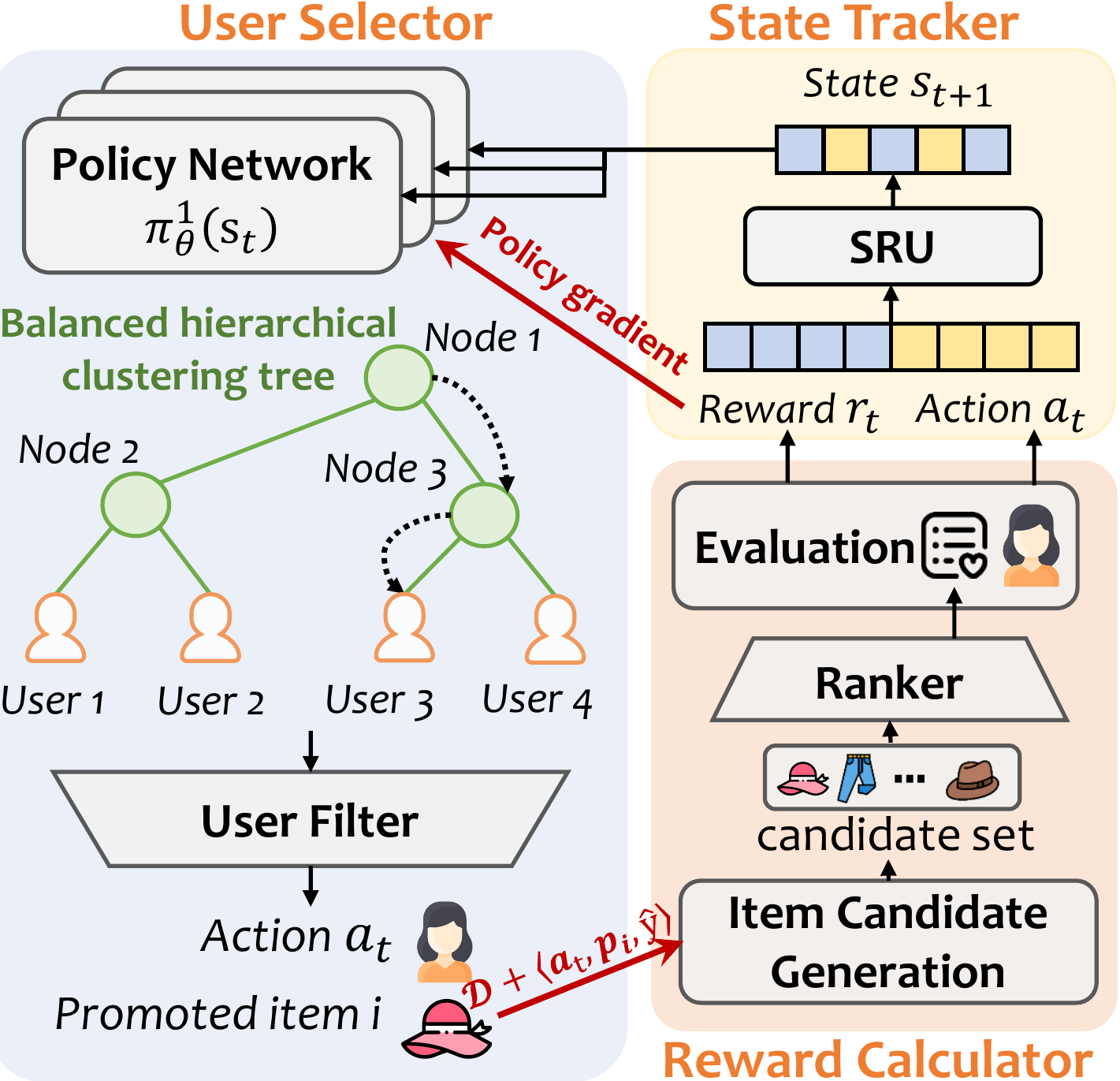}
\centering
\caption{The framework of SMILE.} \label{overall}
\vspace{-0.3cm}
\end{figure}

Figure~\ref{overall} illustrates the framework of our proposed model SMILE, which contains three modules: A state tracker that provides the state vector $\mathbf{s}_t$ based on previous decisions and rewards, a user selector that outputs the selected trial user $a_t$, and a reward calculator that returns a reward signal $r_t$ measuring the effect of the free trial on the recommender system. 
In what follows, we will elaborate on the three modules in detail.

\subsection{State Representation}
\begin{figure}[t]
\includegraphics[width=0.8\linewidth]{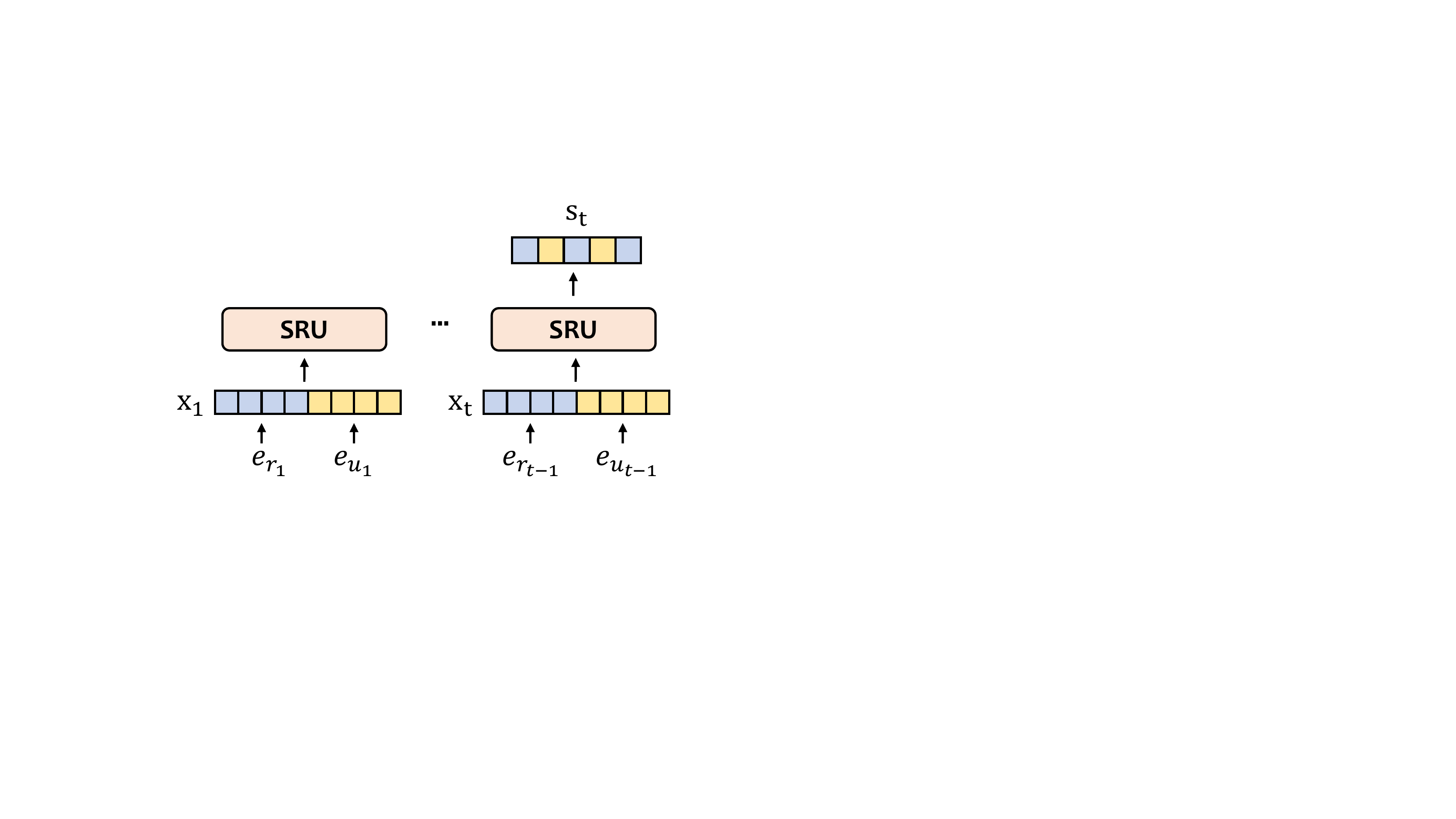}
\centering
\caption{The model for generating the state.} \label{state}
\vspace{-0.3cm}
\end{figure}

The state is designed to understand item preference in each round of the selection. Figure~\ref{state} illustrates the model for generating the state. We adopt a simple recurrent unit (SRU) \cite{lei2018training}, a RNN model that simplifies the computation and exposes more parallelism, to learn the hidden representations.
To integrate historical interactions and feedback information, SRU takes the selected users and the corresponding rewards as input and encodes them into a low-dimensional state vector $\mathbf{s}_i\in \mathbb{R}^{d_s}$. Specially, we set the initial state $\mathbf{s}_1$ as promoted item profiles vector $\mathbf{e}_i$. It can be learned in an end-to-end manner or pre-trained by supervised learning models such as matrix factorization (MF). For the convenience of implementation, we take the pre-trained item matrix to compute $\mathbf{e}_i$ in this part.

Assuming the model is learning the $t$-th state vector $\mathbf{s}_t$, it takes a sequence of previous selected user embeddings $\{{\mathbf{e}_{u_1},\mathbf{e}_{u_2},\cdots,\mathbf{e}_{u_{t-1}}}\}$ and their corresponding reward vectors $\{{\mathbf{e}_{r_1},\mathbf{e}_{r_2},\cdots,\mathbf{e}_{r_{t-1}}}\}$  before timestep $t$ as input.  
Each user is mapped to an embedding vector $\mathbf{e}_{u_i}\in \mathbb{R}^{d_a}$ which is the $i$-th row of user embedding matrix $\mathbf{U}$. $\mathbf{U} \in \mathbb{R}^{\left |u  \right |  \times d_a}$ is pre-trained by MF \cite{koren2009matrix,10.1007/978-3-030-18590-9_72,9665244} and is fixed in the RL process. Equation~\ref{mf} shows the objective function of MF, where $\left |u  \right | $ is the number of users and $\left |i  \right | $ is the number of items. $\mathbf{V} \in \mathbb{R}^{{\left |i  \right | } \times d_a}$ represents the item embedding matrix.  $\mathbf{Y}\in \mathbb{R}^{\left |u  \right |  \times \left |i  \right | }$ denotes the user-item interaction matrix, where $y_{ui}=y$ if the user $u$ rated item $i$ as $y$, otherwise $y_{ui}=0$.

\begin{equation}
\label{mf}
\min _{U \in \mathbb{R}^{{\left |u  \right | } \times d_a}, V \in \mathbb{R}^{{\left |i  \right | } \times d_a}}\left\|Y-U V^{T}\right\|_{F}^{2}.
\end{equation}

Simultaneously, the reward value from $r_{min}$ to $r_{max}$ is linearly mapped into a $h$-dimensional one-hot vector $\mathbf{e}_{r_i}\in \mathbb{R}^{d_h}$ as Equation~\ref{normalize}.
Assuming that the range of reward values is $(r_{min}, r_{max}]$, we firstly normalize each reward value $r$  
to range $(0,h]$ and then utilize the $one\_hot(i, h)$ function to output a $h$-dimensional vector, where the value of the $i$-th element is one and the others are set to zero.

\begin{equation}
\label{normalize}
\begin{aligned}
&{\mathbf{e}_{r_i} }=\operatorname{one\_hot}\left(\hat{r}, h\right), \\
&{\hat{\mathbf{r}}}=h- \left \lfloor\frac{h \times(r_{max}-r)}{r_{max}-r_{min}} \right \rfloor.
\end{aligned}
\end{equation}

To retain richer semantic information, we concatenate the user embedding vector $\mathbf{e}_{u_t}$ and one-hot reward vector $\mathbf{e}_{r_t}$ into $\mathbf{x}_{t} = (\mathbf{e}_{u_t},\mathbf{e}_{r_t})$.
Next, we take $\mathbf{x}_t$ as the input of SRU to learn state representation, and the update function of a SRU cell is defined as
\begin{equation}
\label{sru}
\begin{aligned} \tilde{\mathbf{x}}_{t} &=\mathbf{W} \mathbf{x}_{t}, \\ \mathbf{f}_{t} &=\sigma\left(\mathbf{W}_{f} \mathbf{x}_{t}+\mathbf{b}_{f}\right), \\ \mathbf{g}_{t} &=\sigma\left(\mathbf{W}_{g} \mathbf{x}_{t}+\mathbf{b}_{g}\right), \\ \mathbf{c}_{t} &=\mathbf{f}_{t} \odot \mathbf{c}_{t-1}+\left(1-\mathbf{f}_{t}\right) \odot \tilde{\mathbf{x}}_{t}, \\ \mathbf{h}_{t} &=\mathbf{g}_{t} \odot g\left(\mathbf{c}_{t}\right)+\left(1-\mathbf{g}_{t}\right) \odot \mathbf{x}_{t}, \end{aligned}
\end{equation}
where $\mathbf{x}_t$ denotes the input vector, $\mathbf{f}_t$ and $\mathbf{g}_t$ denote the forget gate and reset gate respectively, $\mathbf{c}_t$ and $\mathbf{h}_t$ indicate the internal state and output state, and $\odot$ is the elementwise product operator. The final hidden state $\mathbf{h}_t$ is the representation of current state $\mathbf{s}_t$, which is then fed into policy network.

\subsection{Architecture of User Selector}
To find the best adopters, we propose a two-stage selecting module called \textit{user selector}. It first utilizes the balanced hierarchical clustering tree structure to select the initial trial user and then leverages a user filter to sort out users further. We will separately introduce the two parts in the following.

\subsubsection{Balanced Hierarchical Clustering over Users}
As mentioned earlier, most RL-based models suffer problems from operating on the large discrete action space which makes the training inefficient and ineffective. In other words, the model has to explore a large discrete action space to find the target adopters to earn positive rewards which makes the time complexity of making a decision linear to the size of the action space. Inspired by the work of Chen \emph{et al.} \cite{chen2019large}, we reformulate the user action space by building a balanced hierarchical clustering tree $\mathcal{T}$ to achieve high effectiveness.
As shown in the \textit{user selector} module in Fig.~\ref{overall}, each leaf node is mapped to a particular user, and each non-leaf node is associated with a policy network. The process of selecting an appropriate user is regarded as a top-down moving from the root to a leaf node.

For the convenience of presentation and implementation, we employ the simple and popular divisive approach to build up the tree $\mathcal{T}$. In this approach, the original data points (i.e., the representation of users) are divided into several clusters, and each cluster is divided into smaller sub-clusters. To make the tree balance, for each node, the difference between heights of its sub-trees is at most one, and the number of child nodes for each non-leaf node is $c$ except for the nodes on the second-to-last layer, whose numbers of child nodes are at most $c$ — considering that the number of users is insufficient to support constructing a perfect $c$-ary tree. 
The value of $c$ is calculated by the whole user set $\mathcal{U}$ and the tree depth $d$, which is defined as follows:

\begin{equation}
\label{c}
    c=\left \lceil \left |\mathcal{U} \right | ^{\frac{1}{d} }  \right \rceil,
\end{equation}
where $\lceil x \rceil $ returns the smallest integer no less than $x$.

Next, we employ the PCA-based clustering algorithm to perform the balanced hierarchical clustering over users, which takes a group of user vectors ${\{\mathbf{e}_{u_1},\mathbf{e}_{u_2}, \cdots, \mathbf{e}_{u_m}\}}$ and the number of child nodes $c$ as inputs. Specially, the input user vector $\mathbf{e}_{u_i} \in \mathbb{R}^{d_a}$ is the $i$-th row of the user embedding matrix $\mathbf{U}$. Then, these vectors are divided into $c$ balanced clusters. By repeatedly applying the clustering algorithm until each sub-cluster is associated with only one user, a balanced clustering tree is successfully constructed.

Taking a scenario with four users for illustration, the \textit{user selector} module in Fig.~\ref{overall} shows the constructed balanced clustering tree with the tree depth $d$ set to two. On the tree $\mathcal{T}$, each leaf node ($user_1 \sim user_4$) represents a user $u \in \mathcal{U}$ and each non-leaf node ($node_1 \sim node_3$) has an independent policy network $\pi_{\theta}$.
To begin with, a path $\mathcal{P}$ starts at the root node ($node_1$) which takes the aforementioned state $\mathbf{s}_t$ as input and outputs a probability distribution over $c$ child nodes. And the node ($node_3$) with the maximum probability will be extended to the path.
Then, the path $\mathcal{P}$ keeps extending until reaching a leaf node then the corresponding user ($user3$) is the selected user.

Accordingly, getting a trial user at timestep $t$ is the process of generating a path ${\mathcal{P}_t}=\{c_1,c_2, \cdots, c_d\}$ from the root node to a leaf node. It consists of $d$ (i.e., the number of layers in the tree) choices, and each choice is represented as an integer between one and $c$ (i.e., the maximum number of child of each node). Given the state, the probability of action at timestep $t$ is

\begin{equation}
\label{pro}
\pi_{\theta}\left(a_{t} \mid \mathbf{s}_{t}\right)=\prod_{i=1}^{d} \pi_{\theta_{i}}\left(c_{i} \mid \mathbf{s}_{t}\right),
\end{equation}
where $ \pi_{\theta_{i}}\left(c_{i} \mid \mathbf{s}_{t}\right)$ is the probability of making each choice in the corresponding policy network from root to the chosen action which is computed in Equation~\ref{softmax}. Our goal is to optimize all the policy network set $\mathbf{\pi_{\theta}}=\{\pi_{\theta_1}, \pi_{\theta_2},{\cdots}, \pi_{\theta_{\mathcal{Q}}}\}$, where $\mathcal{Q}$ denoting the number of non-leaf nodes of the tree which is computed by $\mathcal{Q}=\frac{c^d-1}{c-1}$.

\subsubsection{Filtering out Inappropriate Users}
By constantly leveraging the tree structure, we can get a set of initial trial users. 
Unfortunately, not every individual is fond of the free trial items. Indiscriminately providing items to all the primary selected users may hurt customer experiences and reduce their intention of final purchasing. Meanwhile, the platform may receive some low ratings or negative comments, resulting in a negative impact on sales. Consequently, it is crucial to select users who favor the promoted items and are more inclined to give high ratings.

To this end, we design a user filter module as shown in the \textit{user filter} module in Fig.~\ref{overall} to mimic user preference towards target items based on MF. Without loss of generality, we regard the ratings higher than 3.5 as positive ratings (notice that the highest rating is five) and take this criterion to eliminate inappropriate users. 
In this way, we can get the filtered users (i.e., action $a_t$) who are interested in the promoted items and are more likely to give high ratings. And we regard them as the ultimately selected adopters.

\subsection{Building Reward Function}
With the help of the user selector module, trial adopter $a_t$ is obtained successfully. Then, we adopt the free trial activity to collect immediate user feedback $\left \langle u,p_{i},\hat{y} \right \rangle $ and append the triplet to the original dataset $\mathcal{D}$, where $\hat{y}$ denotes the predicted rating between the adopter $u$ and the promoted item $i$ computed by Equation~\ref{mf}.
In this way, $\mathcal{D}$ embraces the newly created interactions as well as historical interactions concurrently.

Next, we develop a measurable indicator that takes the dataset $\mathcal{D}$ as input and outputs the free trial's effect on the recommender system. Intuitively, real-time sales of promoted items is a favorable indicator reflecting whether the products sell well. But it is impractical to train our model online with real users to capture sales changes because three reasons: (1) For the model, the increment in sales is slow and usually takes weeks to collect sufficient data to make the assessment statistically significant; (2) for users, interacting with a half-baked system can hurt experiences; (3) for the platform, collecting real-time user feedback requires expensive engineering and logistic overhead \cite{jagerman2018opensearch,jagerman2019people,DBLP:journals/corr/abs-2101-09459}. 

In this paper, we introduce the concept of Page View (PV) \cite{li2019fair,garcin2014offline}, which is an available evaluation indicator to measure items’ exposure within a certain period on the recommender system. We define our reward value as the average exposure of the target promoted item set $I_p$, which is represented as follows: 

\begin{equation}
\label{reward}
\operatorname{\mathcal{R}(\mathbf{s}_t,a_t)}=\sum_{u \in \mathcal{U}}\frac{\left|L_{u} \cap I_{p}\right|}
{\left | I_p \right | } ,
\end{equation}
where $\mathcal{U}$ denotes the whole user set, $L_u$ represents the recommended $K$ items to user $u$, and $I_p$ is the target item set to be promoted. 

As shown in the \textit{reward calculator} module in Fig.~\ref{overall}, the recommended results $L_u$ are generated by the \textit{item candidate generation} and the \textit{ranker} modules \cite{song2020poisonrec}. Specifically, the item candidate generation \cite{covington2016deep} module selects hundreds of items from the entire item corpus to construct a candidate set $C_u$ for each user $u$.
The ranker \cite{he2017neural,hidasi2015session,wang2019neural,wu2019session} is responsible for ranking the items in $C_u$ based on Bayesian Personalized Ranking (BPR) \cite{rendle2012bpr} algorithm, which can estimate user's preference score on items. Then the $K$ items with the highest scores will be recommended in the final recommendations list $L_u$. 
Therefore, the reward value $r_t$ given state $\mathbf{s}_t$ and action $a_t$ is calculated by Equation~\ref{reward} and is regarded as the signal guiding the whole optimization process.

\begin{algorithm}[t]
	\caption{The procedure of SMILE}
	\LinesNumbered 
	\label{alg:algorithm}
	\KwIn{Episode length $n$, Tree depth $d$, Promoted item set $I_p$, Original data $\mathcal{D}$, Reward function $\mathcal{R}$, User set $\mathcal{U}$ with representations}
	\KwOut{Model parameters $\theta$}
	\DontPrintSemicolon
	Calculate the number of child nodes $c$ and non-leaf nodes $\mathcal{Q}$ \\
	Construct a balanced clustering tree $\mathcal{T}$ with $c$ child nodes\\
	\For {$j=1$ to $\mathcal{Q}$}{
		initialize $\theta_j \leftarrow $ random values\\
	}
    \Repeat {converged} {
    	\For {$t=1$ to $n$}{
    		Sample ${\mathcal{P}}_t=\{c_1,c_2, \cdots, c_d\}$\\
    		Map $p_t$ to a user $a_t$ after passing user filter\\
    			\For {$i=1$ to $|I_p|$}{
    		$\mathcal{D}=\mathcal{D} + <a_t,p_{i},y>$\\
    		}
    		$r_t=\mathcal{R}(\mathbf{s}_t,a_t)$\\
    		\If{$t<n$}{
    		Calculate $\mathbf{s}_{t+1}$ by state tracker\\ 
    		}
    	}
	    Get $\mathcal{M}=(\mathbf{s}_1,a_1,r_1,\cdots,\mathbf{s}_n,a_n,r_n)$\\
       \For{$t=1$ to $n$} {
          Update $\theta$ according to Equation~\ref{update}\\
        }
    }
    \textbf{return} $\theta$\\
\end{algorithm}

\subsection{Model Optimization with Policy Gradient}
As mentioned earlier, we utilize a policy network to learn the strategy of choosing the best subclass at each non-leaf node given the current state. Our main idea is to find the best adopters that can maximize the exposure of promoted items. As illustrated in Fig.~\ref{overall}, for one thing, the output of the reward calculator $r_t$ will enter the state tracker module to learn the next state vector $\mathbf{s}_{t+1}$; for another thing, $r_t$ will be used to train the policy network set $\mathbf{\pi_{\theta}}=\{\pi_{\theta_1}, \pi_{\theta_2},{\cdots}, \pi_{\theta_{\mathcal{Q}}}\}$ in the balanced hierarchical clustering tree in user selector module. 

We utilize the most commonly used policy gradient methods REINFORCE algorithm \cite{williams1992simple} to train our model. The objective is to maximize the expected discounted cumulative rewards, i.e.,
\begin{equation}
\label{rein}
J\left(\pi_{\theta}\right)=\mathbb{E}_{\pi_{\theta}}\left[\sum_{t=0}^{n-1} \gamma^{t} r_{t}\left(\mathbf{s}_t, a_{t}\right)\right],
\end{equation}
where $\gamma$ is the discount factor, $\eta$ is the learning rate, and $r_{t}$ denotes target items’ exposure within a certain period on recommender system after the free trial process. 
We can update the parameter ${\theta}$ by Equation~\ref{update}:

\begin{equation}
\label{update}
\begin{aligned}
&{Q}^{\pi_{\theta}}\left(\mathbf{s}_t, a_{t}\right)=\sum_{i=t}^{n} \gamma^{i-t} r_{i},\\
&\Delta \theta= \nabla_{\theta} \log \pi_{\theta}\left(a_{t} \mid \mathbf{s}_t\right) {Q}^{\pi_{\theta}}\left(\mathbf{s}_t, a_{t}\right),\\
&\theta=\theta+\eta \Delta \theta,\\
\end{aligned}
\end{equation}
where $ \pi_{\theta}(a_t \mid \mathbf{s}_t) $ denotes the probability of taking action $a_t$ given state $\mathbf{s}_t$; $Q^{\pi_{\theta}}(s, a)$ is the cumulative rewards.  
The overall training process is shown in Algorithm~\ref{alg:algorithm}.

\subsection{Complexity Analysis}
In this section, we discuss the complexity of SMILE from time complexity and space complexity. As every policy network is implemented as a full-connected layer, we consider both the time and the space complexity of each policy network as $\mathcal{O}(c)$. 

\textbf{Time complexity.}
Considering the process of selecting one user, the state vector will via $d$ policy networks and each time needs to choose from at most $c$ options. Therefore, the time complexity of selecting one user is $\mathcal{O}(d\times c)\simeq \mathcal{O}(d\times \left | A^{\frac{1}{d} } \right | )$. As we usually set the value of tree depth $d$ to a small number, our tree structure method can significantly reduce the time complexity compared with other RL-based methods whose time complexity is $\mathcal{O}(|A|)$, where $A$ denotes the action space.

\textbf{Space complexity.} The space complexity mainly comes from two parts: the number of policy network (i.e., the number of non-leaf nodes) and its optional range (i.e., the child nodes). We firstly calculate the number of policy network $\mathcal{Q}$. The space complexity of the SMILE is  $\mathcal{O}(\mathcal{Q}\times c) \simeq  \mathcal{O}(\frac{c^{d}-1}{c-1} \times c)\simeq \mathcal{O}(\left | A \right | )$, which is equal to other RL-based models.

%% file: LaTeX/04experiment.tex
\section{Experiments and Results}

We conducted extensive experiments on three public datasets to justify our model's superiority and reveal the reasons for its effectiveness. 

In this section, we first introduce the statistics of the datasets and present five baselines whose selection strategies are based on user behavior patterns. Besides, we design two evaluation metrics for the reward of the selection models and use two more metrics for the influence of the selection over the recommendation system. We also specify the experiment setup for the evaluation. 

Specifically, we will answer the following research questions to unfold the experiments.

\textbf{RQ1}: What is the influence of varying the number of trial users on the exposure effect?

\textbf{RQ2}: Compared with the static free trial user selection policy, how does our model perform?

\textbf{RQ3}: What are the benefits of the tree structure and the impact of tree depths in our model? 

\textbf{RQ4}: How does the free trial process influence the effectiveness of the recommender system?

\subsection{Datasets} We conducted experiments on three public datasets as follows, and the statistical information of these datasets is shown in Table~\ref{dataset}.
\begin{itemize}
    \item \textbf{Movielens100K\footnote{http://grouplens.org/datasets/movielens/100k/}}: Movielens100K \cite{harper2015movielens} consists of 100,000 movie ratings of 943 users for 1,682 movies. 
    \item  \textbf{Movielens1M\footnote{https://grouplens.org/datasets/movielens/1m/}}: Movielens1M \cite{harper2015movielens} contains one million anonymous ratings of 3,900 movies by 6,040 users.
    \item  \textbf{Ciao\footnote{http://www.cse.msu.edu/ tangjili/trust.html}}: Ciao \cite{tang2012mtrust} is collected from a real-world social media website. From the originally dataset, we filter out users who rated less than three items and items that received less than three ratings, which leaves us 6,626 users, 15,048 items, and 161,813 ratings.
\end{itemize}

\subsection{Baselines}
Since there is a lack of study investigating the problem of finding the best trial adopters for item promotion, we take five static policies based on user behavior patterns as our baselines:

\begin{itemize}
    \item \textbf{Random}: selecting trial adopters at random.
    \item \textbf{Activity}: ranking users according to their activity (i.e., the number of user transaction volume) and taking the most active users as the trial adopters.
    \item \textbf{Inactivity}: contrary to activity strategy, the inactivity strategy takes the least active users as the trial adopters.
    \item \textbf{HighRating}: ranking users according to their historical ratings and selecting users who prefer to score high ratings as the trial adopters.
    \item \textbf{LowRating}: contrary to highRating strategy, lowRating strategy selects users who prefer to score low ratings as the trial adopters.
\end{itemize}

\begin{table}[ht]
\caption{The statistics of datasets.}
\label{dataset}
\begin{tabular}{c|cccl}
\hline
\textbf{DataSet}       & \textbf{\#Users} & \textbf{\#Items} & \textbf{\#Ratings} & \textbf{Density} \\ \hline
\textbf{Movielens100K} & 943              & 1,682            & 100,000            & 6.30\%           \\
\textbf{Movielens1M}   & 6,040            & 3,900            & 1,000,209          & 4.25\%           \\
\textbf{Ciao}          & 7,935            & 16,200           & 171,465            & 0.13\%           \\ \hline
\end{tabular}
\end{table}

\subsection{Evaluation Metrics}
We design two metrics for evaluating the rewards of selection models. As the RL-based methods aim to gain the optimal long-run rewards, we use the average reward ($Avg\_reward$) over each selection episode for each promoted item as one evaluation metric. Besides, we adopt the maximum reward ($Max\_reward$) value to measure the best performance of selection strategies quickly.

Besides, we employ two widely adopted metrics $Precision@k$ and $Recall@k$ \cite{herlocker2004evaluating} with $k= 10$ to measure the free trial influence over recommender system. $Precision@k$ is the proportion of recommended items in the top-$k$ set that are relevant. $Recall@k$ is the fraction of relevant items that have been retrieved in the top-$k$ relevant items.

\subsection{Experimental Setup}
\subsubsection{Simulating User Preference}
As mentioned before, not everyone favors the promoted items, so we design a user filter module to simulate adopters' preferences on them and filter those who are not interested. Empirically, the promoted item set $I_p$ always lacks exposure opportunities, i.e., possessing subtle interaction data. To overcome the obstacle of accurately mimicking users' predilections on promoted items, we expressly set $I_p$ as the popular items with considerable interaction data which helps improve prediction accuracy. Next, to imitate the low exposure feature of promoted items, we delete part of their transactions and reconstruct a new dataset. Finally, the user filter module is trained on the original dataset, ensuring the prediction precision between adopters and promoted items. 
Taking the movielens100k dataset as an example, we first set the promoted items $I_p$ as the top 1\% popular items and train our prediction model (i.e., MF) on the original dataset. Then we delete their transactions until the original 5\% interaction data is retained. In this way, we can utilize the entire dataset to make predictions and the processed dataset is used in subsequent experiments.

\subsubsection{Generating Candidates}
\begin{figure}[ht]
    \centering
    \includegraphics[width=0.8\linewidth]{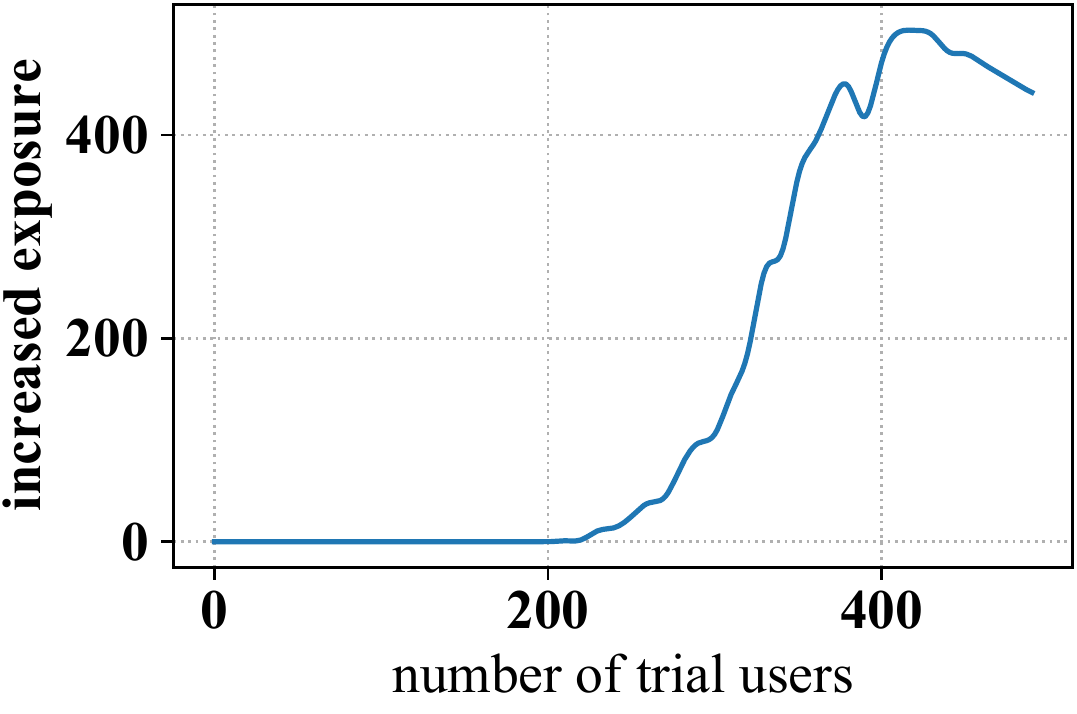}
    \caption{Influence of the increased adopters' number on exposure effect.}
    \label{fig:inv1}
    \vspace{-0.2cm}
\end{figure}

To explore the influence of adopters' number on the exposure effect, we select an increasing number of adopters linearly by random metric. Figure~\ref{fig:inv1} reflects that not a more significant number of adopters achieve higher increased exposures. The curve keeps rising at first, and after achieving a peak, it gradually drops. Considering more adopters requires more free items, which is none other than a considerable expense. Therefore, we must control the number of adopters, and it is urgent to set a suitable selection policy to winnow users and achieve high exposure.

Taking the movielens1M dataset as an example, a candidate item set $c_u$  is made up by the promoted set $I_p$ and the 10\% other items selected randomly. For the recommendation results, we assume that each user only views $K$ items and defines the items with the highest estimated preference scores in ranker as $L_u$. Therefore, there will be a high reward if target items frequently appear in users' recommendation results $L_u$.


\begin{table*}[ht]
\centering
\caption{Overall selection performance comparison.}\label{overall_reward}
\begin{tabular}{ccccccc}
\hline
\textbf{DataSet}    & \multicolumn{2}{c}{\textbf{Movielens100K}}  & \multicolumn{2}{c}{\textbf{Movielens1M}}    & \multicolumn{2}{c}{\textbf{Ciao}}           \\
\textbf{Metric}     & \textbf{Avg\_reward} & \textbf{Max\_reward} & \textbf{Avg\_reward} & \textbf{Max\_reward} & \textbf{Avg\_reward} & \textbf{Max\_reward} \\ \hline
\textbf{Random}     & 4.72                 & 36                   & 10.72                & 40                   & 32.88                    & 47                    \\
\textbf{Activity}   & 0.14                 & 7                    & 9.78                 & 69                   & 22.42                    & 35                    \\
\textbf{Inactivity} & 11.57                & 54                   & 15.46                & 43                   & 37.5                    & 51                    \\
\textbf{HighRating} & 8.54                 & 55                   & 13.40                & 54                   & 34.75                    & 45                    \\
\textbf{LowRating}  & 3.93                 & 24                   & 6.80                 & 27                   & 33.8                    & 40                    \\ \hline
\textbf{SMILE}       & \textbf{138.3}       & \textbf{213}         & \textbf{55.7}        & \textbf{89}          & \textbf{62}           & \textbf{69}           \\ \hline
\end{tabular}
\end{table*}

\subsubsection{Implementation Details}
In our experiments, we aim to select 5\% adopters among all users, which is the same as the episode length. Once a user $u$ is sampled in each episode, it will be removed from the available users; thus, no repeated users occur in an episode.
For the balanced hierarchical clustering tree, we set the tree depth $d$ to two which can achieve the best performance.
In the optimization process, we set the discount factor $\gamma$ to 0.9 and optimize all models with the Adam optimizer. 

\subsection{Investigation on The Number of Adopters (\textbf{RQ1})}
In real-world marketing scenarios, the platform will avoid selecting too many trial users considering the limited money and time. To investigate the influence of trial user number on the exposure effect, we conduct two experiments on the Movielens100K dataset.

Figure~\ref{fig:inv1} shows the change of total increased exposure when we incrementally raise the number of adopters by random selection policy. Surprisingly, they are not always positively correlated, i.e., more adopters do not correspond to more exposure opportunities, which is somewhat counterintuitive. In the beginning, the increased exposure is zero because of the tiny number of adopters. Later, with more adopters selected, the exposure curve begins to rise, indicating that the promotion effect has been achieved. However, the curve gradually decreases after reaching the peak, suggesting that superfluous adopters will reduce the marketing effect. It may be because that the newly added connections violate the authentic user preferences distribution.
As a consequence, more adopters require high costs and time but may fail to achieve better performance. It confirms the necessity of exploring a suitable free trial user selection policy that aims to achieve high exposure at the lowest expense.

\begin{figure}[ht]
    \centering
    \includegraphics[width=0.8\linewidth]{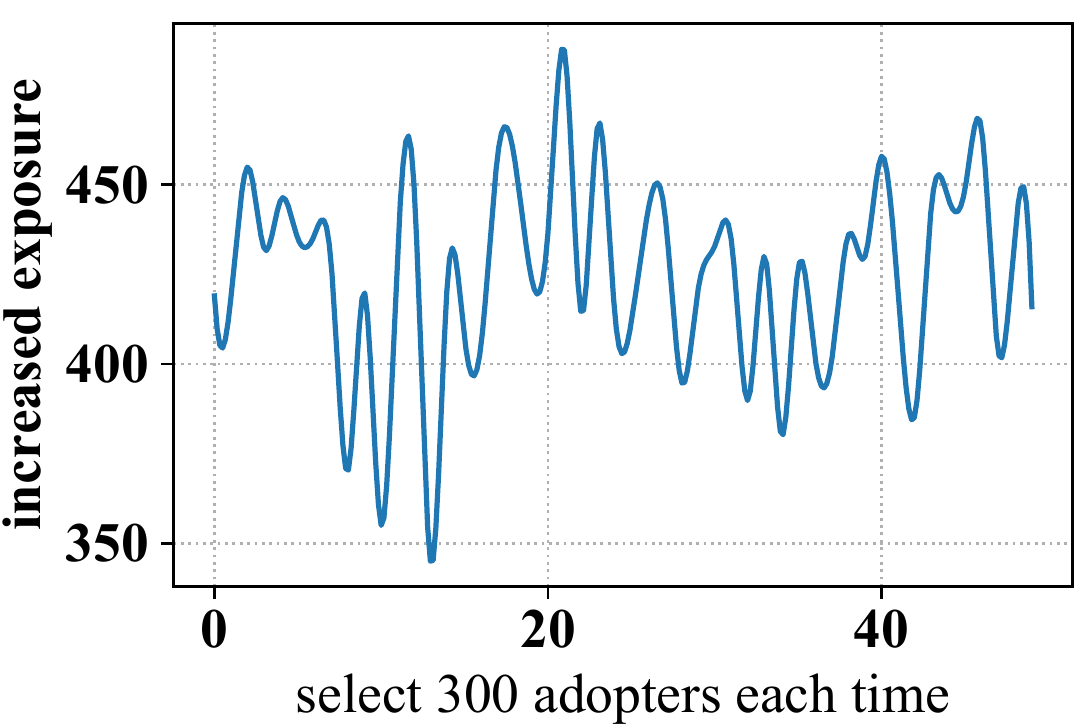}
    \caption{Influence of the same number of adopters on exposure effect.}
    \label{fig:inv2}
    \vspace{-0.2cm}
\end{figure}

Figure~\ref{fig:inv2} illustrates the different increased exposure with the same number (three hundred) of adopters every time. Interestingly, the result is not stable as we wish, fluctuating around four hundred. This inconsistency could be attributed to the inaccurate performance of the recommendation algorithm. Therefore, it is rational to record the maximum reward value as an evaluation metric to reflect the potential maximum exposure effect.

\subsection{Overall Selection Performance Comparison (\textbf{RQ2})}
To validate the superiority of SMILE, we conduct experiments on three datasets. The summarized results are presented in Table \ref{overall_reward}. We highlight the best results of all models in boldface. According to the results, we note the following observations:

1) The inactive selection strategy gets a higher average reward than the active one. It might be because the inactive users hold much fewer interactions thus are more sensitive to new interactions and more likely to affect the recommendation system. We can also observe that users who prefer to score high ratings achieve higher reward value users prefer low ratings. The reason is associated with the user filter module, which tends to drop users favoring low ratings. 
2) The maximum rewards of the five baseline selection strategies are almost similar, indicting the randomness and instability of selecting users by their attributes. On the whole, it is hard to find a fixed selection strategy from the baselines for the best performance. We argue that these rigid and stationary methods cannot adjust to the flexible reality scene. 
3) Among these methods, our proposed SMILE framework achieves the best performance in both metrics. Especially in the Movielens100K dataset, the average reward is far greater than the maximum reward of baselines, reflecting its significant advantages in small datasets. The main reasons are threefold. First, it adopts RL technology for long-run planning and dynamic adaptation, which is absent in other baselines. Second, the hierarchical clustering tends to cluster similar users in the same subtree, which incorporates additional user similarity information into our model. Third, the hierarchical tree-structured can ease the training process to some degree.

\begin{table*}[ht]
\centering
\caption{The free trial influences over recommender systems.}\label{influence}
\begin{tabular}{ccccccc}
\hline
\textbf{DataSet}     & \multicolumn{2}{c}{\textbf{Movielens100K}} & \multicolumn{2}{c}{\textbf{Movielens1M}}   & \multicolumn{2}{c}{\textbf{Ciao}}          \\
\textbf{Metric}      & \textbf{Precision@10} & \textbf{Recall@10} & \textbf{Precision@10} & \textbf{Recall@10} & \textbf{Precision@10} & \textbf{Recall@10} \\ \hline
\textbf{Original}    & 0.2194                & 0.0505             & 0.1032                & 0.0347             & 0.0326                & 0.0167             \\
\textbf{SMILE}     & 0.2364                & 0.0538             & 0.1066                & 0.0378             & 0.0395                & 0.0198             \\ \hline
\textbf{Improvement} & \textbf{7.75\%}       & \textbf{6.53\%}    & \textbf{3.29\%}       & \textbf{8.93\%}    & \textbf{21.4\%}       & \textbf{18.7\%}    \\ \hline
\end{tabular}
\end{table*}

\subsection{Benefits of Hierarchical Clustering Tree (\textbf{RQ3})}

\begin{figure}[ht]
    \centering
    \includegraphics[width=0.9\linewidth]{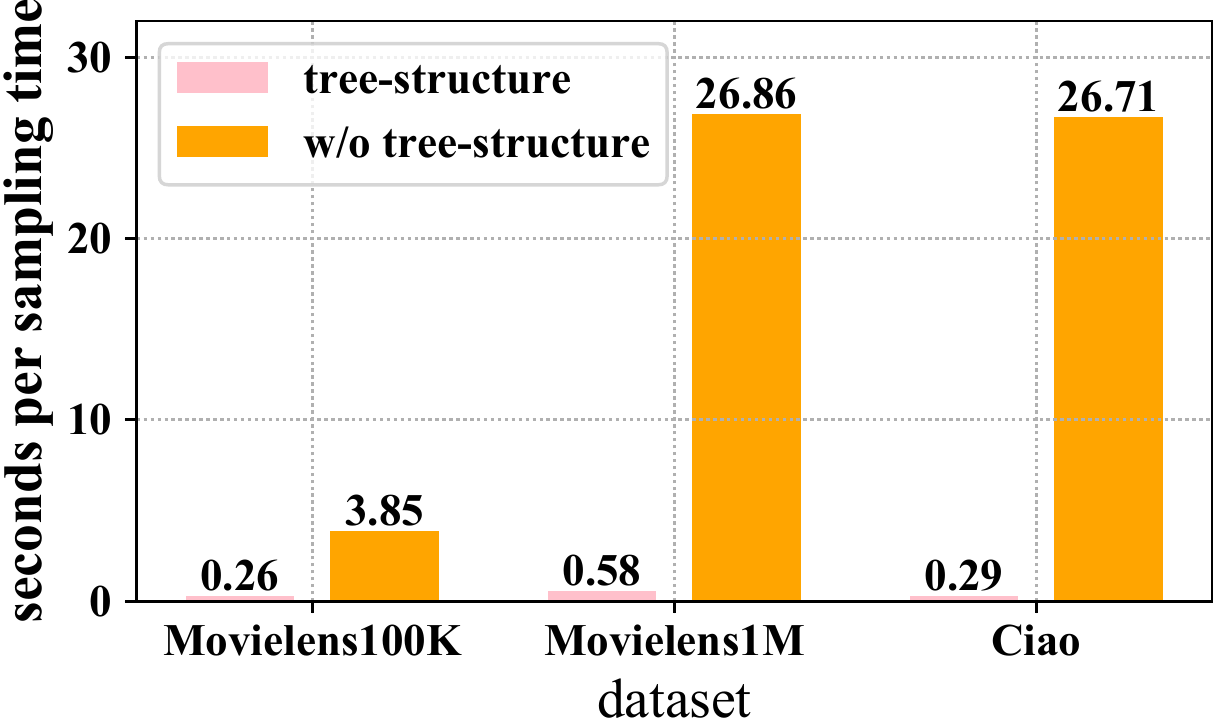}
    \caption{Influence of tree structure.}
    \label{fig:bar}
    \vspace{-0.2cm}
\end{figure}

In the user selector module, we conduct a tree-structured decomposition and adopt a certain number of policy networks with simple architectures. To investigate its feasibility and efficiency, we conduct two experiments.
First, we compare the running time in the sampling stage between the model with the hierarchical clustering tree structure and the model without a tree structure (i.e., only preserves one policy network, which takes a state as input and gives the policy possibility distribution on all users) on three datasets. To make the comparison fairly, all the experiments are conducted on the same machine with i7-6850K CPU @ 3.60GHz. As shown in Fig.~\ref{fig:bar}, we can easily observe that our hierarchical clustering tree structure takes the shortest running time when sampling an action.

\begin{figure}[ht]
    \centering
    \includegraphics[width=0.9\linewidth]{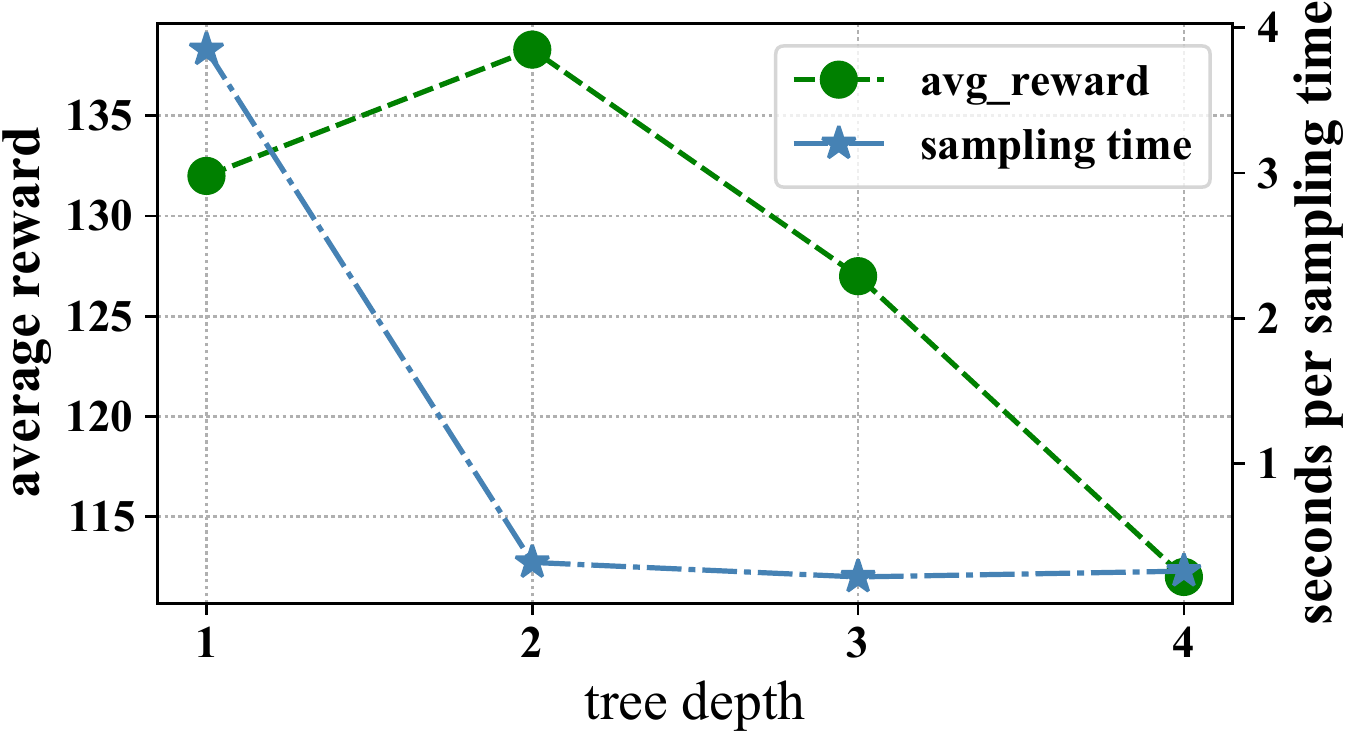}
    \caption{Results under different tree depth.}
    \label{fig:line}
    \vspace{-0.2cm}
\end{figure}

Next, we set depth from one to four to further explore how different tree depth influences the efficiency and performance of SMILE on the Movielens100K dataset. In particular, the model with tree depth set to one is equivalent to without a tree structure. The green curve in Fig.~\ref{fig:line} representing the sampling time shows that the tree structure model significantly improves efficiency. The blue curve presents the performance under different tree depths, from which we can notice that the model with the tree depth set to two reaches the peak performance, while other tree depths cause a slight performance drop. Therefore, setting the depth of the tree to two can significantly reduce the time complexity and provide better performance.

\subsection{Free Trial Influences over RS (\textbf{RQ4})}
To simulate the trial process, we establish a triplet $\left \langle u,p_i,\hat{y} \right \rangle $ denoting the new interactions between adopter $u$ and promoted target item $p_i$. However, will the implementation change the actual data distribution and then degrade the performance of the recommendation algorithm? To answer this question, we conduct an experiment to explore the effects of the free trial influences over the BPR recommendation algorithm and explain the applicability of our proposed SMILE model.

As shown in Table \ref{influence}, instead of reducing its performance, SMILE can improve the accuracy of recommendations. Especially on the Ciao dataset, it achieves an increment by 21.4\% and 18.7\% in $Precision@10$ and $Recall@10$, respectively. The improvement of the recommender system is equivalent to the increase of users' probability of purchasing products from the recommendation list, which is none other than a fantastic signal indicating the expansion of the global sale on the platform.

This increment may be due to the following two reasons. First, the selected trial users are all interested in promoted items; hence no original data distribution is changed (i.e., the newly added interaction data is consistent with users' historical preferences). Second, more interactions will provide richer information for model learning that alleviates the data-sparse problem and further improve the performance.  

%% file: LaTeX/05conclusion.tex
\section{Conclusion and Future Work}
In this paper, we systematically analyze and formulate the problem of selecting suitable adopters to increase item exposure in the scenario of the recommender system. We propose a novel free trial user selection model named SMILE, which consists of three modules: A state tracker that provides a state vector based on previous decisions and rewards, a user selector that produces selected adopter based on hierarchical clustering over user action space, and a reward calculator that evaluates the selection performance. At last, we utilize policy gradient to update our model. Experiments conducted on three public datasets demonstrate that our proposed SMILE framework can achieve better performance with higher efficiency.


In the future, we seek to tackle the adopter selection problem in the social network environment as the message diffusion in the social graph is a significant factor influencing the promotion effect. We also plan to introduce offline reinforcement learning in our scenario, i.e., learning a debiased user model based on offline data and providing reward value to train our reinforcement learning policy. More powerful reinforcement learning algorithms such as Proximal Policy Optimization \cite{ppo} and Deep Deterministic Policy Gradient \cite{ddpg} will also be taken into consideration in our future work.


%% file: LaTeX/06ack.tex
\ifCLASSOPTIONcompsoc
  \section*{Acknowledgments}
\else
  \section*{Acknowledgment}
\fi

This work was supported in part by the National Key Research and Development Program of China (2020YFB1712900), the National Natural Science Foundation of China (62176028), and the Natural Science Foundation of Chongqing (cstc2020jcyj-msxmX0690).